\documentclass[twocolumn,secnumarabic,amssymb, nobibnotes, aps, prd]{revtex4}

\usepackage{amsmath,amssymb}
\usepackage{graphicx}
\usepackage{setspace}
\usepackage{flushend}
\usepackage{hyperref}

\begin{document}
	\title{Perturbation level interacting dark energy model and its consequence on late-time cosmological parameters}
	\author{Mahnaz Asghari}
	\email{std\_m.asghari@khu.ac.ir}
	\affiliation{Department of Astronomy and High Energy Physics, Faculty of Physics, Kharazmi University, Tehran 15719-14911, Iran}
	\author{Shahram Khosravi}
	\email{khosravi\_ sh@khu.ac.ir}
	\affiliation{Department of Astronomy and High Energy Physics, Faculty of Physics, Kharazmi University, Tehran 15719-14911, Iran}
	\author{Amir Mollazadeh}
	\email{amirmollazadeh@khu.ac.ir}	
	\affiliation{Department of Astronomy and High Energy Physics, Faculty of Physics, Kharazmi University, Tehran 15719-14911, Iran}

	\begin{abstract}
	In the present paper, we study the capability of interacting dark energy model with pure momentum transfer in the dark sector to reconcile tensions between low redshift observations and cosmic microwave background (CMB) results. This class of interacting model with pure momentum exchange introduces modifications to the standard model in the level of perturbation. We investigate the model by comparing to observational data, including integrated Sachs-Wolfe-galaxy cross-correlation, galaxy power spectrum, $f \sigma_8$, and CMB data. It is shown that this model can alleviate the observed tension between local and global measurements of $\sigma_8$. According to our results, the best fit value of $\sigma_8$ for interacting model is $0.700$, which is lower than the one for $\Lambda$CDM model and also is consistent with low redshift observations. Furthermore, we perform a forecast analysis to find the constraints on parameters of the interacting model from future experiments. 	
	\end{abstract}
	\maketitle
	\section{Introduction} \label{sec1}
	According to distance measurements of type Ia supernovae, it is generally accepted that the Universe is currently experiencing a phase of accelerated expansion \cite{1538-3881-116-3-1009,0004-637X-517-2-565}. In the standard Einstein gravity, a dark energy component that has negative pressure is responsible for this late-time acceleration. There are more cosmological observations such as cosmic microwave background (CMB) anisotropy \cite{2014A&A...571A...1P,2014A&A...571A..16P,2016A&A...594A..13P}, large scale structure (LSS) \cite{PhysRevD.69.103501,1538-3881-128-1-502,1538-3881-129-3-1755}, and integrated Sachs-Wolfe (ISW) effect \cite{1967ApJ...147...73S,1538-4357-683-2-L99,2004Natur.427...45B} that provide strong evidence for the existence of dark energy.
	
	The most convincing candidate for dark energy is a cosmological constant that gives rise to the concordance ${\Lambda}$CDM model for the Universe. Although the ${\Lambda}$CDM model is in reasonable agreement with observational data \cite{doi:10.1111/j.1365-2966.2009.15812.x,0067-0049-208-2-19}, there are two important theoretical difficulties in this model: the cosmological constant problem \cite{RevModPhys.61.1}, and the coincidence problem \cite{Velten2014,PhysRevLett.85.4434,PhysRevLett.85.5276,ZLATEV1999570,Jimenez:2008er}. Additionally, the low redshift data such as late-time determinations of the Hubble constant \cite{0004-637X-826-1-56,0004-637X-775-1-13} and local measurements of $\sigma_8$ from LSS \cite{doi:10.1046/j.1365-8711.2003.06550.x} are in conflict with the Planck CMB data \cite{2016A&A...594A..13P}. These discrepancies might imply an inadequacy of the standard cosmological model and therefore provide a motivation to search for alternative cosmological models to describe the evolution of Universe.
	
	Considering dark matter and dark energy as dominant components in late-time Universe, a nongravitational coupling in the dark sector may seem justifiable from the theoretical point of view \cite{PhysRevD.79.123506,1475-7516-2010-05-009,PhysRevD.84.023010}. Moreover, interacting dark energy models would alleviate problems in ${\Lambda}$CDM model by modifying the background as well as perturbative evolution of Universe \cite{PhysRevD.79.063518,MohseniSadjadi2010,1475-7516-2009-07-027,PhysRevD.89.103531,PhysRevD.92.123537,1475-7516-2016-04-014,PhysRevD.94.043518,1475-7516-2017-01-028,PhysRevD.94.123511,PhysRevD.95.043513,PhysRevD.95.023515,PhysRevD.96.103511,A_2019}.
	
	Interaction between dark matter and dark energy affects the growth of cosmic structure, so it is worth to use LSS information and redshift space distortion (RSD) data for measuring the coupling strength between the dark components. Moreover, interaction in the dark sector would influence CMB anisotropies in large scales as ISW effect \cite{PhysRevD.80.103514}. The late-time ISW effect, which corresponds to the effect of a time varying gravitational potential energy on the CMB photons on large scales, seems a suitable probe for constraining cosmological models concerning the dark energy \cite{2016A&A...594A..21P}. Because of the cosmic variance and the low amplitude of the ISW signal, it is advisable to detect this effect by the cross-correlation between the CMB temperature anisotropy and the large scale structure. The cross-correlation signal can be used to constrain alternative dark energy models \cite{1475-7516-2016-09-003}.
	
	In this paper, we consider a pure momentum exchange interacting model, which makes a modification to the standard model in the perturbation level. Interacting models with pure momentum transfer in the dark sector were first investigated in \cite{simpson,pourtsidou}. In our model, the interaction term is proportional to the relative velocity of dark energy and dark matter, which is a generalization of the interacting dark energy model studied in \cite{A_2019}, in which interaction only affects perturbations and leaves the background evolution similar to the ${\Lambda}$CDM model. So, it is expected that our model solves the tension in structure growth measurements, especially $2-3\sigma$ tension in $\sigma_8$ \cite{2016A&A...594A..13P,10.1093/mnras/stw2665}, while leaving the tensions corresponding to the background. Accordingly, we try to constrain the coupling between dark components by observational data, including ISW-galaxy cross-correlation, galaxy power spectrum, $f \sigma_8$, and CMB data. Furthermore, we explore the validity of interacting dark energy model in comparison with the standard cosmological model. We also perform a Fisher-based forecast in order to constrain the interacting dark energy model with future data.
	
	The outline of the paper is as follows: in Sec. \ref{sec2}, we describe the interacting dark energy model as well as the effect of interaction on CMB and matter power spectra. Section \ref{sec3} is devoted to observational probes used to constrain the interacting model. In Sec. \ref{sec4}, we describe the methods used in the analysis and also present the results containing best fit values of cosmological parameters of the studied interacting model. In Sec. \ref{sec5}, we perform forecast analysis to find constraints on the interaction in the dark sector by using a Fisher matrix analysis. We conclude in Sec. \ref{sec6}.
	\section{The interacting dark energy model} \label{sec2}
	We assume a late-time interacting dark energy model, in which the dark sector components are perfect fluids with the following energy-momentum tensor:
	\begin{equation} \label{eq1}
	T^\mu_{\nu\,\mathrm{(A)}}=(\rho_{\mathrm{(A)}}+p_{\mathrm{(A)}})\,u^\mu_{\mathrm{(A)}}\,u_{\nu\,\mathrm{(A)}}+\delta^\mu_\nu\,p_{\mathrm{(A)}} ,
	\end{equation}
	where $\rho_{\mathrm{(A)}}$, $p_{\mathrm{(A)}}$, and $u^\mu_{\mathrm{(A)}}$ represent the energy density, pressure, and the four velocity of component $A$ in the Universe, respectively.
	In addition, considering linear perturbations of the perturbed Friedmann-Lema\^itre-Robertson-Walker metric in conformal Newtonian gauge gives
	\begin{equation} \label{eq2}
	\mathrm{d} s^2=a^2 \, \left(-(1+2\mathrm{\Psi})\mathrm{d} \tau^2+(1-2\mathrm{\Phi})\mathrm{d} \vec{x}^2\right) ,
	\end{equation}
	with the conformal time $\tau$. $\mathrm{\Psi}$ and $\mathrm{\Phi}$ are the Newtonian potentials, and in the absence of anisotropic stress, we have $\mathrm{\Psi}=\mathrm{\Phi}$. So, the four velocity of component $A$ takes the form
	\begin{equation} \label{eq3}
	u^\mu_\mathrm{(A)}=\frac{1}{a}\,(1-\mathrm{\Phi},v^i_{\mathrm{(A)}}) .
	\end{equation}
	It is known that the total energy-momentum tensor in the Universe is conserved,
	\begin{equation} \label{eq4}
	\nabla_{\mu}\,T^\mu_{\nu\,\mathrm{(tot)}}=0 ,
	\end{equation}
	however, for interacting dark energy components, the energy-momentum tensors are not conserved separately, so that
	\begin{align}
	&\nabla_{\mu}\,T^\mu_{\nu\,\mathrm{(DM)}}=Q_{\nu\,\mathrm{(DM)}} ,  \label{eq5} \\
	&\nabla_{\mu}\,T^\mu_{\nu\,\mathrm{(DE)}}=Q_{\nu\,\mathrm{(DE)}} ,   \label{eq6}
	\end{align}
	where $Q_{\nu \mathrm{(A)}}$ is the interaction term corresponding to component \textit{A}. Considering the fact that the nature of dark matter and dark energy is unknown, it is possible to study interacting dark energy models in a phenomenological approach. In general, $Q^\mu_{\mathrm{(A)}}$ can be written as
	\begin{equation} \label{eq7}
	Q^\mu_{\mathrm{(A)}}=Q_{\mathrm{(A)}}\,u^\mu+F^\mu_{\mathrm{(A)}} ,
	\end{equation}
	where $Q_{\mathrm{(A)}}=\bar{Q}_{\mathrm{(A)}}+\delta Q_{\mathrm{(A)}}$ is the energy density transfer rate, $u^\mu$ is the total four velocity, and $F^\mu_{\mathrm{(A)}}=\frac{1}{a}\,(0,\partial^i f_{\mathrm{(A)}})$ is the momentum density transfer rate (with $f_{\mathrm{(A)}}$ the momentum transfer potential).	
	Conservation of the total energy-momentum tensor would impose
	\begin{align}
	&\sum_{A}\,Q^\mu_{\mathrm{(A)}}=0 , \label{eq8} \\
	&\sum_{A}\,Q_{\mathrm{(A)}}=0 , \label{eq9} \\
	&\sum_{A}\,f_{\mathrm{(A)}}=0 . \label{eq10}
	\end{align}
	Thus, according to (\ref{eq8}), we have
	\begin{equation} \label{eq11}
	Q^\mu_{\mathrm{(DM)}}=-Q^\mu_{\mathrm{(DE)}}=Q^\mu .
	\end{equation}
	
	Following the approach of \cite{Koshelev2011}, $Q^\mu_{\mathrm{(DM)}}$ is assumed to be proportional to a linear combination of dark sector velocities,
	\begin{equation} \label{eq12}
	Q^\mu_{\mathrm{(DM)}}=Q\,(\gamma_{\mathrm{(DM)}}\,u^\mu_{\mathrm{(DM)}}+\gamma_{\mathrm{(DE)}}\,u^\mu_{\mathrm{(DE)}}) .
	\end{equation}
	Considering Eq. (\ref{eq7}), we can write
	\begin{align} 
	& Q\,(\gamma_{\mathrm{(DM)}}\,u^\mu_{\mathrm{(DM)}}+\gamma_{\mathrm{(DE)}}\,u^\mu_{\mathrm{(DE)}})=Q_{\mathrm{(DM)}}\,u^\mu+F^\mu_{\mathrm{(DM)}} , \label{eq13} \\
	& \to F^\mu_{\mathrm{(DM)}}=Q\,(\gamma_{\mathrm{(DM)}}\,u^\mu_{\mathrm{(DM)}}+\gamma_{\mathrm{(DE)}}\,u^\mu_{\mathrm{(DE)}})-Q_{\mathrm{(DM)}}\,u^\mu . \label{eq14}
	\end{align}
	Regarding the fact that $F^0_{\mathrm{(DM)}}=0$, it is possible to write
	\begin{equation} \label{eq15}
	 F^0_{\mathrm{(DM)}}=Q\,(\gamma_{\mathrm{(DM)}}\,u^0_{\mathrm{(DM)}}+\gamma_{\mathrm{(DE)}}\,u^0_{\mathrm{(DE)}})-Q_{\mathrm{(DM)}}\,u^0=0 ,
	\end{equation}
	 which gives
    \begin{equation*}
	Q_{\mathrm{(DM)}}=Q\,(\gamma_{\mathrm{(DM)}}+\gamma_{\mathrm{(DE)}})=-Q_{\mathrm{(DE)}} .
	\end{equation*}
	In a special case, we choose
	\begin{equation*}
	\gamma_{\mathrm{(DE)}}=-\gamma_{\mathrm{(DM)}}=\gamma ,
	\end{equation*}
	for which Eq. (\ref{eq12}) takes the form
	\begin{equation} \label{i1}
	Q^\mu_{\mathrm{(DM)}}=Q\,\gamma\,(u^\mu_{\mathrm{(DE)}}-u^\mu_{\mathrm{(DM)}}) .
	\end{equation}
	Also, it is found that
	\begin{align}
	& Q_{\mathrm{(DM)}}=0 , \\
	& F^\mu_{\mathrm{(DM)}}=Q\,\gamma\,(u^\mu_{\mathrm{(DE)}}-u^\mu_{\mathrm{(DM)}}) . \label{i3}
	\end{align}
	Hence, it is easy to see that the energy transfer vanishes in the dark sector, and there is only momentum exchange between dark matter and dark energy. The temporal and spatial components of the interaction term (to linear order of perturbations) would be
	\begin{align}
	& Q^0=0 , \label{eq16} \\
	& Q^i=\frac{1}{a}\,\bar{Q}\,\gamma\,(v^i_{\mathrm{(DE)}}-v^i_{\mathrm{(DM)}}) . \label{eq17}
	\end{align}
	Following \cite{Koshelev2011}, $\bar{Q}$ can be written as \cite{Mangano-10.1142}
	\begin{equation} \label{eq18}
	\bar{Q}\propto M^{5-4\alpha-4\beta}\,\bar{\rho}^{\alpha}_{\mathrm{(DM)}} \,\bar{\rho}^{\beta}_{\mathrm{(DE)}} ,
	\end{equation}
	where $\alpha$ and $\beta$ are constants, and $M$ is a parameter with dimension of energy. By choosing $\alpha=\beta=\frac{1}{2}$, and $M=\mathcal{H}$ (with $\mathcal{H}$ the conformal Hubble parameter), the spatial components of interaction term take the form
	\begin{equation} \label{eq19}
	Q^i=\frac{1}{a}\,\gamma\,\mathcal{H}\,\sqrt{\bar{\rho}_{\mathrm{(DM)}}\,\bar{\rho}_{\mathrm{(DE)}}}\,(v^i_{\mathrm{(DE)}}-v^i_{\mathrm{(DM)}}) ,
	\end{equation}
	where $\gamma$ is the dimensionless coupling constant.
		
	Now, we can derive the covariant conservation equations (\ref{eq5}) and (\ref{eq6}) in background and perturbative levels, applying (\ref{eq16}) and (\ref{eq19}) as interaction terms. The background continuity equations are
	\begin{align}
	&\bar{\rho}'_{\mathrm{(DM)}}+3\,\mathcal{H}\,\bar{\rho}_{\mathrm{(DM)}}=0 , \label{eq20} \\
	&\bar{\rho}'_{\mathrm{(DE)}}+3\,\mathcal{H}\,\bar{\rho}_{\mathrm{(DE)}}\,(1+w_{\mathrm{(DE)}})=0 , \label{eq21}
	\end{align}
	where a prime indicates derivative with respect to the conformal time. It is evident that the background level is similar to the ${\Lambda}$CDM model. The linearized continuity and Euler equations in conformal Newtonian gauge are given by
	\begin{align}
	\delta'_{\mathrm{(DM)}}=-\theta_{\mathrm{(DM)}}+3\,\mathrm{\Phi}' , \label{eq22}
	\end{align}
	\begin{align}
	\theta'_{\mathrm{(DM)}}&=-\mathcal{H}\,\theta_{\mathrm{(DM)}}+k^2\,\mathrm{\Phi} \nonumber \\
	&+\gamma\,a\,\mathcal{H}\,\sqrt{\frac{\bar{\rho}_{\mathrm{(DE)}}}{\bar{\rho}_{\mathrm{(DM)}}}}\,(\theta_{\mathrm{(DE)}}-\theta_{\mathrm{(DM)}}) , \label{eq23}
	\end{align}
	\begin{align}
	\delta'_{\mathrm{(DE)}}&=-3\,\mathcal{H}\,\delta_{\mathrm{(DE)}}(c^2_{s\mathrm{(DE)}}-w_{\mathrm{(DE)}}) \nonumber \\
	&-\theta_{\mathrm{(DE)}}(1+w_{\mathrm{(DE)}})\big(1+9\frac{\mathcal{H}^2}{k^2}(c^2_{s\mathrm{(DE)}}-c^2_{a\mathrm{(DE)}})\big) \nonumber \\
	&+3\,(1+w_{\mathrm{(DE)}})\,\mathrm{\Phi}' , \label{eq24}
	\end{align}
	\begin{align}
	\theta'_{\mathrm{(DE)}}&=\theta_{\mathrm{(DE)}}\mathcal{H}\big(-1+3\,w_{\mathrm{(DE)}}+3(c^2_{s\mathrm{(DE)}}-c^2_{a\mathrm{(DE)}})\big) \nonumber \\
	&+k^2\,\mathrm{\Phi}+\frac{k^2\,c^2_{s\mathrm{(DE)}}}{1+w_{\mathrm{(DE)}}}\,\delta_{\mathrm{(DE)}} \nonumber \\
	&-\gamma\,\frac{a\mathcal{H}}{1+w_{\mathrm{(DE)}}}\,\sqrt{\frac{\bar{\rho}_{\mathrm{(DM)}}}{\bar{\rho}_{\mathrm{(DE)}}}}(\theta_{\mathrm{(DE)}}-\theta_{\mathrm{(DM)}}) , \label{eq25}
	\end{align}
	where $\delta_\mathrm{(A)}$ is the density contrast, and $\theta_\mathrm{(A)}=i k_i v^i_\mathrm{(A)}$ is the divergence of velocity perturbation of component \textit{A}. We have considered constant dark energy equation of state, and so the adiabatic sound speed of dark energy is defined as $c^2_{a\mathrm{(DE)}}=w_{\mathrm{(DE)}}$.
	
	The Einstein field equations would not be directly affected by interaction, so the linearized gravitational field equations in conformal Newtonian gauge are
	\begin{align}
	& k^2\,\Phi+3\,\mathcal{H}^2\,\Phi+3\,\mathcal{H}\,\Phi'=-4\pi G\,a^2\,\sum_{j} \delta \rho_{(j)} , \label{e1} \\
	& k^2\,\Phi'+\mathcal{H}\,k^2\,\Phi=4\pi G\,a^2\,\sum_{j} (\bar{\rho}_{(j)}+\bar{p}_{(j)})\,\theta_{(j)} , \label{e2} \\
	& \Phi''+3\,\mathcal{H}\,\Phi'+(2\mathcal{H}'+\mathcal{H}^2)\,\Phi=4\pi G\,a^2\,\sum_{j} \delta p_{(j)} , \label{e3}
	\end{align}
	where \textit{j} indicates all components in the Universe (containing photons, neutrinos, baryons, dark matter, and dark energy).
	
	In order to see the influence of interaction on CMB temperature power spectrum, we modify the Cosmic Linear Anisotropy Solving System (CLASS) code \cite{1475-7516-2011-07-034} according to the interacting dark energy model described in Eqs. (\ref{eq22})--(\ref{eq25}). For this purpose, we use adiabatic initial conditions defined in the CLASS code, and besides the Planck 2015 results \cite{2016A&A...594A..13P} for cosmological parameters, we assume $w_{\mathrm{(DE)}}=-0.99$ (to avoid divergences in dark energy perturbation equations) and $c^2_{s\mathrm{(DE)}}=1$.
	
\begin{figure*}
		\centering
		\includegraphics[width=8.5cm]{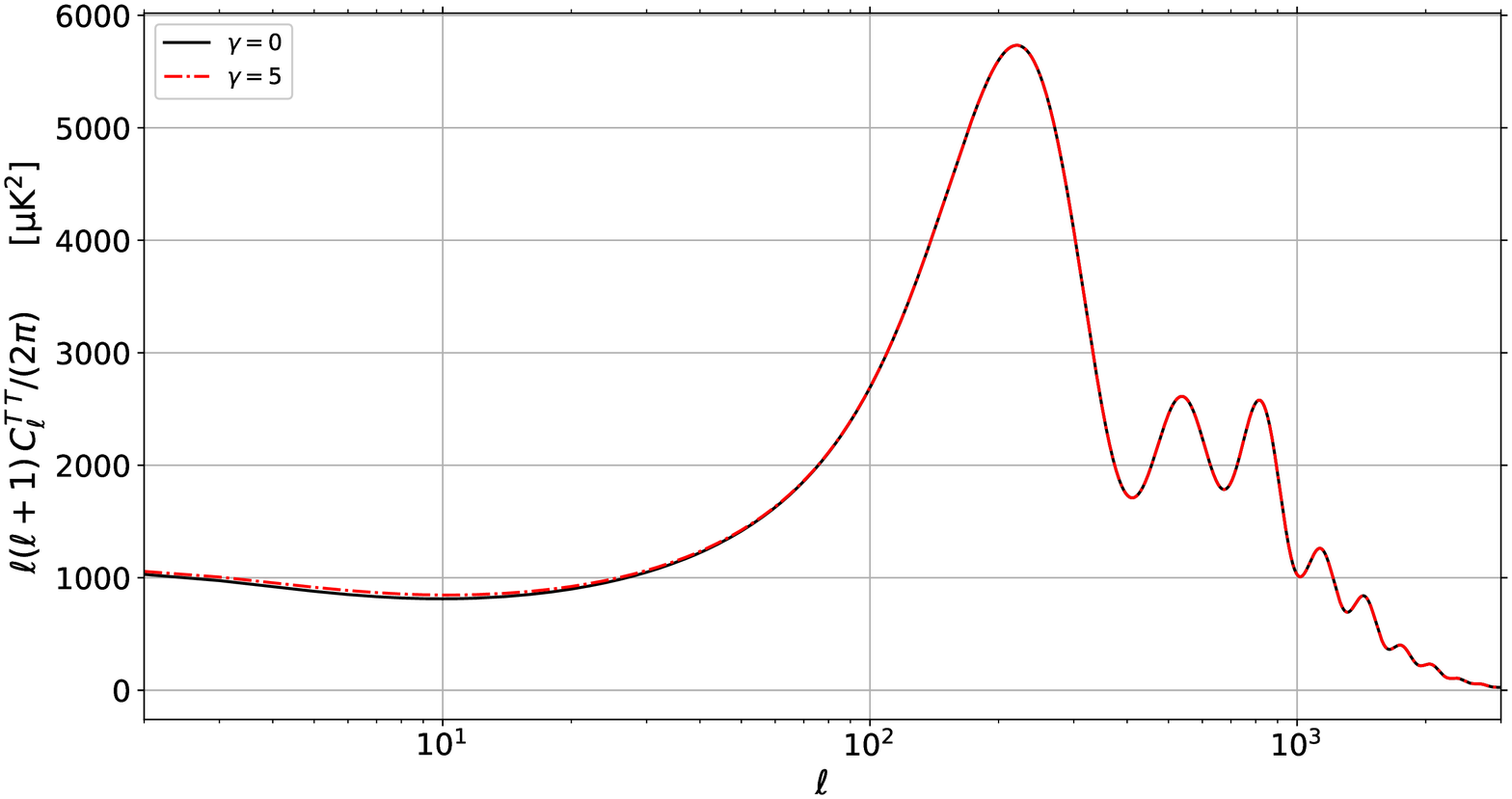}
		\includegraphics[width=8.5cm]{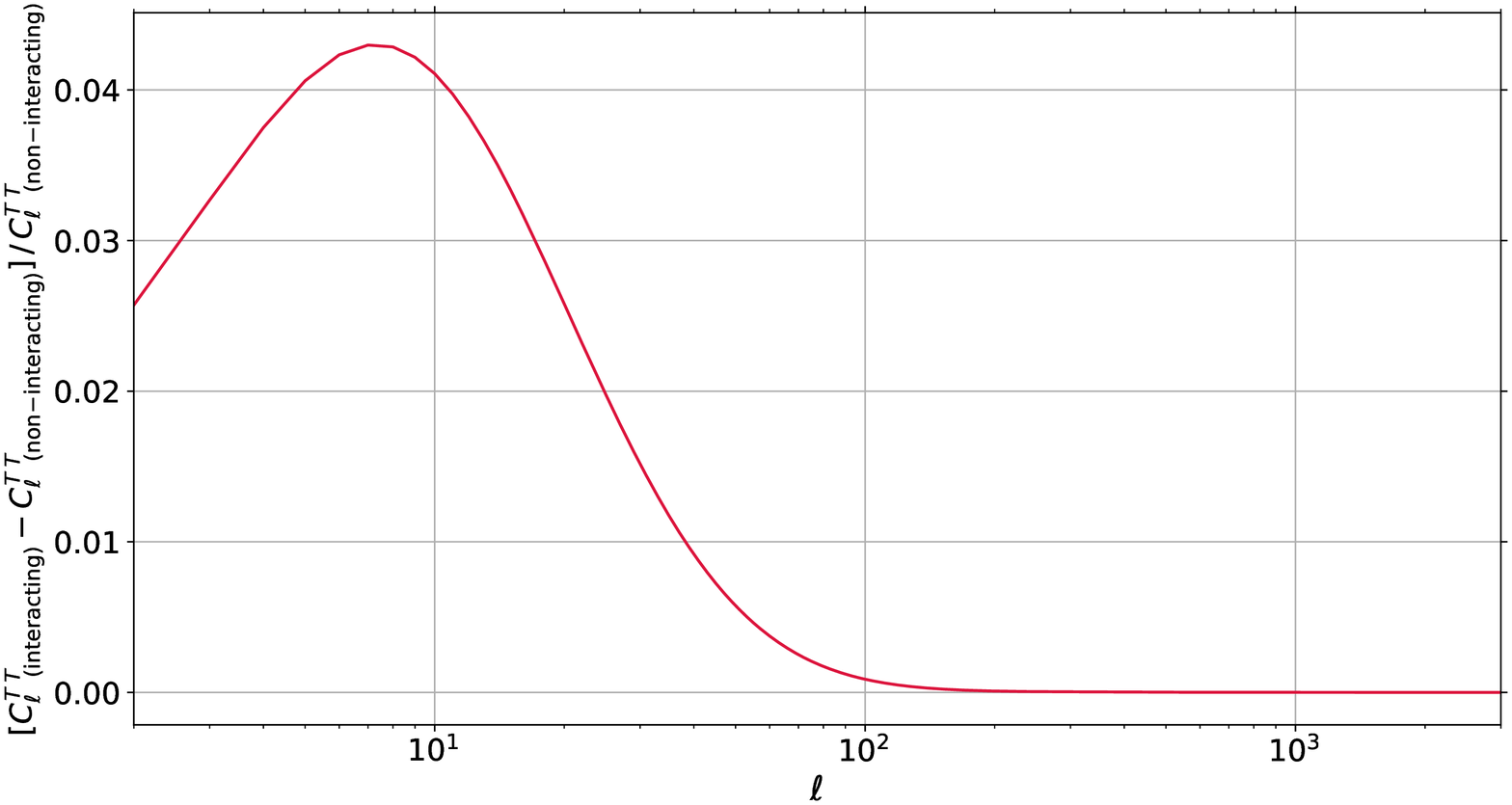}
		\caption{Left: the \textit{TT} component of CMB power spectrum, considering noninteracting model ($\gamma=0$) and interacting model with $\gamma=5$. Right: relative difference diagram of CMB anisotropy power spectrum, with $\gamma=5$.}
		\label{fig1}
	\end{figure*}

\begin{figure}[h!] 
		\centering
		\includegraphics[width=8.5cm]{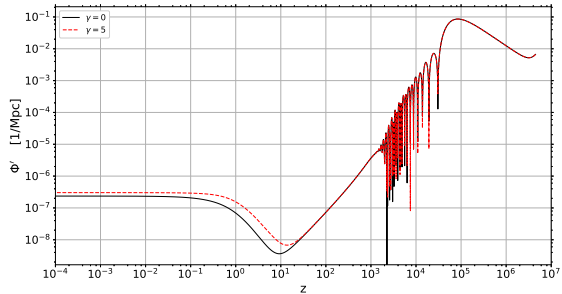}
		\caption{$\mathrm{\Phi}'$ in terms of redshift for noninteracting model ($\gamma=0$) and interacting model with $\gamma=5$.}
		\label{fig2}
	\end{figure}

	Figure \ref{fig1} shows the CMB temperature anisotropy spectrum for interacting and noninteracting models along with the relative difference diagram of two models.
	According to the fact that interaction would not appear in background level, there is no change in location and height of acoustic peaks in the CMB power spectrum. However, in large scales, we can see an increase in CMB anisotropies caused by the ISW effect. Considering Eq. (\ref{eq26}), the temperature anisotropy due to ISW effect is related to $\mathrm{\Phi}'$. On the other hand, interaction in the dark sector which modifies perturbative equations would change the potential perturbation $\mathrm{\Phi}$ and its derivative with respect to conformal time. As shown in Fig. \ref{fig2}, interaction in the dark sector would increase $\mathrm{\Phi}'$ compared to noninteracting model and consequently result in enhancement in CMB anisotropy spectrum.
	
\begin{figure}[h!]
		\centering
		\includegraphics[width=8.5cm]{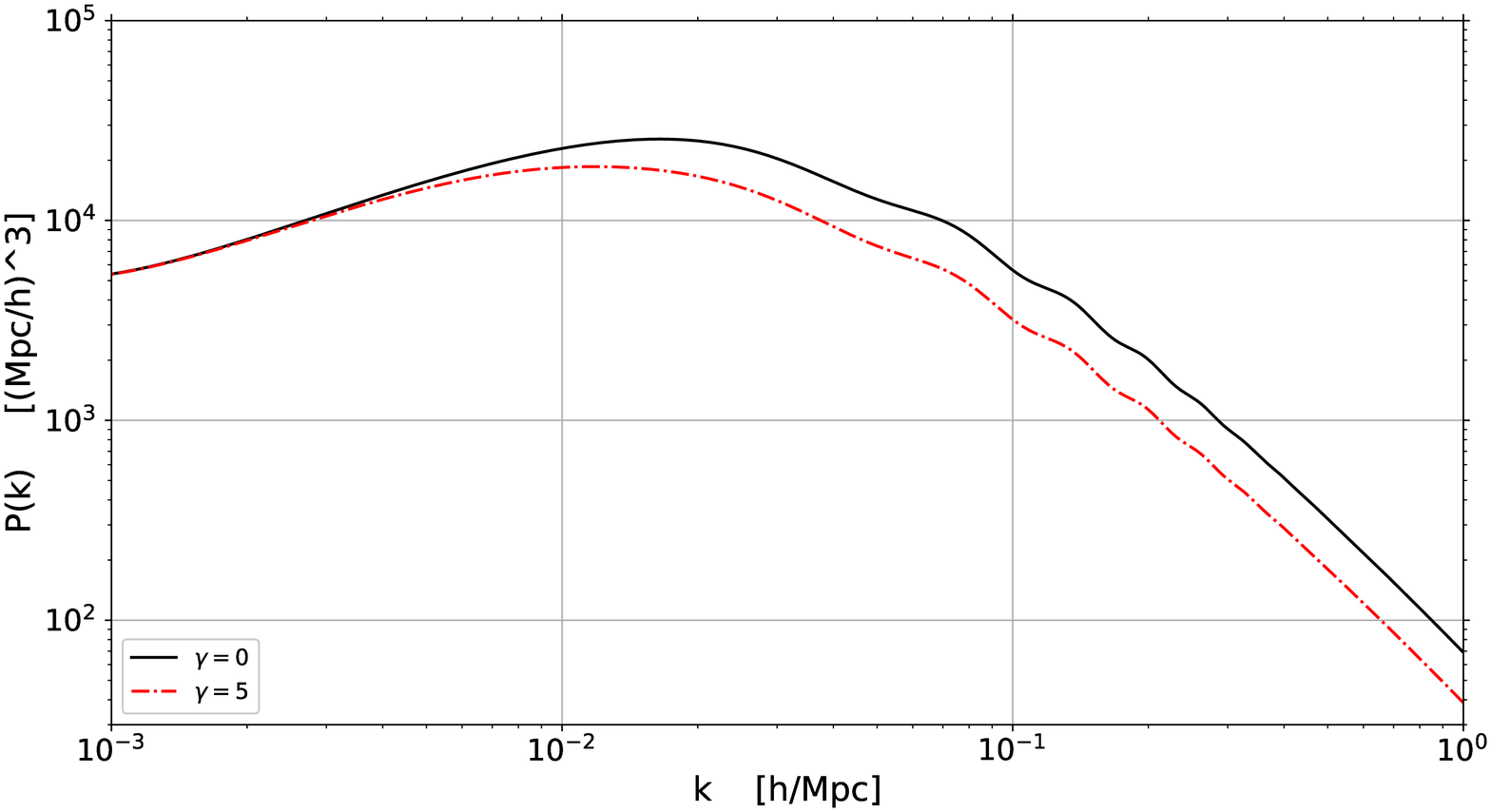}
		\caption{Matter power spectrum diagrams for noninteracting model ($\gamma=0$) and interacting model with $\gamma=5$.}
		\label{fig3}
	\end{figure}

	It is worth to mention that interaction in the dark sector would suppress structure growth as illustrated in matter power spectrum diagrams in Fig. \ref{fig3}, obtained from the modified version of the CLASS code according to Eqs. (\ref{eq22})--(\ref{eq25}).
	According to the evolution equations of the interacting model, interaction appears in perturbation level, and so it is more important in smaller scales. So only sub-Hubble modes would be affected by the interaction, while supper-Hubble modes evolve similarly to ${\Lambda}$CDM model.
	
	In order to explain suppression in matter power spectrum (caused by the interaction), qualitatively, one can consider some theoretical analyses: it is possible to derive the evolution of dark matter density contrast from Eqs. (\ref{eq22})--(\ref{eq25}) [also with using Eqs. (\ref{e1})--(\ref{e3})]. Considering sub-Hubble modes and also neglecting dark energy perturbations compared to dark matter ones (since we have considered $c^2_{s\mathrm{(DE)}}=1$), $\delta_{\mathrm{(DM)}}$ evolves as 
	\begin{align}
	&\delta''_{\mathrm{(DM)}}+\delta'_{\mathrm{(DM)}}\,\mathcal{H}\,(1+\gamma\,a\,\sqrt{\frac{\bar{\rho}_{\mathrm{(DE)}}}{\bar{\rho}_{\mathrm{(DM)}}}}) \nonumber \\
	&-\frac{a^2}{2\,M^2_{\mathrm{pl}}}\,\bar{\rho}_{\mathrm{(DM)}}\,\delta_{\mathrm{(DM)}}=0 . \label{eqdm}
	\end{align} 
	According to Eq. (\ref{eqdm}), the interacting parameter appears in Hubble friction term. So, it can be easily seen that in noninteracting case ($\gamma=0$), dark matter perturbations evolve similar to the ${\Lambda}$CDM model. However, for interacting model with positive values of coupling constant, there is a drag due to dark energy component (with $c^2_{s\mathrm{(DE)}}=1$) on the dark matter, preventing dark matter from clustering and hence suppressing structure growth in the Universe.
	\section{Observational probes} \label{sec3}	
	In this section, we introduce the cosmological observables that are used to constrain the parameters of the interacting dark energy model.
	\subsection{The ISW-galaxy cross-correlation} \label{sec3.1}
	Here, we discuss the cross-correlation between the temperature fluctuations due to the ISW effect and galaxy density contrast. Because of the primordial anisotropies as well as cosmic variance on large scales, the ISW signal is difficult to detect directly. However, it is shown that the signal could be detected through cross-correlation of the CMB with a local tracer of mass \cite{PhysRevLett.76.575,nature02139,PhysRevD.71.123521,PhysRevD.78.043519,PhysRevD.91.083533}.
	 
	 Briefly describing the ISW effect, photons of the CMB travel through the gravitational potential of large scale structures from the last scattering surface to the present day. By entering a gravitational potential due to a structure, photons gain energy and then lose energy when leaving the well again. In the dark energy dominated era, the potential well of a structure becomes shallower as time passes, which results in a net shift in the wavelength of photon, and the total shift in wavelength leads to a secondary anisotropy in the CMB temperature distribution. Time dependence of the gravitational potential introduces an effect on temperature perturbations of CMB photons coming from the direction $\vec{\theta}$ on celestial sphere,
	\begin{equation} \label{eq26}
	{(\frac{\delta T}{T})}_{\mathrm{ISW}}(\vec{\theta})=-\,\int^{\tau_0}_{\tau_i} {\mathrm{d}\tau\,e^{-\tau_\mathrm{op}}\,\psi'(\vec{\theta},\tau)} ,
	\end{equation}
	where the integral is taken from a prerecombination time $\tau_i$ to the present time $\tau_0$, and $\tau_\mathrm{op}$ is the optical depth (which we neglect it). The ISW potential $\psi$ is defined as
	\begin{equation} \label{eq26+}
	\psi(\vec{\theta},\tau)=\mathrm{\Phi}(\vec{\theta},\tau)+\mathrm{\Psi}(\vec{\theta},\tau)=2\,\mathrm{\Phi}(\vec{\theta},\tau) .
	\end{equation}
	Taking Fourier transform and writing in term of redshift give
	\begin{equation} \label{eq30}
	{(\frac{\delta T}{T})}_{\mathrm{ISW}}(\vec{\theta})=\frac{2}{{(2\pi)}^3}\,\int{\mathrm{d}^3 k\int^{z_i}_{0} {\mathrm{d}z\,e^{i\vec{k}.\vec{r}}\,\,\frac{\partial \mathrm{\Phi}(\vec{k},z)}{\partial z}}} .
	\end{equation}
	Also, it is possible to expand the temperature anisotropy due to ISW effect in terms of spherical harmonics,
	\begin{equation} \label{eq31}
	{(\frac{\delta T}{T})}_{\mathrm{ISW}}(\vec{\theta})=\sum_{\ell m}\,a^T_{\ell m}\,Y_{\ell m}(\vec{\theta}) .
	\end{equation}
	On the other hand, the (00) component of Einstein equations in subhorizon regime ($k^2\gg\mathcal{H}^2$) gives
    \begin{equation} \label{eq33}
    k^2\,\mathrm{\Phi} \simeq -\frac{a^2}{2\,M^2_{\mathrm{pl}}}\,(\bar{\rho}_{\mathrm{(DM)}}\,\delta_{\mathrm{(DM)}}+\bar{\rho}_{\mathrm{(DE)}}\,\delta_{\mathrm{(DE)}}) ,
    \end{equation}	
    where we have assumed a late-time universe containing dark matter and dark energy. By introducing parameter $q=1+\frac{\bar{\rho}_{\mathrm{(DE)}}\,\delta_{\mathrm{(DE)}}}{\bar{\rho}_{\mathrm{(DM)}}\,\delta_{\mathrm{(DM)}}}$, Eq. (\ref{eq33}) takes the form
    \begin{align} \label{eq34}
    & k^2\,\mathrm{\Phi} \simeq -\frac{a^2}{2\,M^2_{\mathrm{pl}}}\,\bar{\rho}_{\mathrm{(DM)}}\,\delta_{\mathrm{(DM)}}\,q , \nonumber \\
    & \to \, \mathrm{\Phi}(\vec{k},z)=-\frac{3\,H_0^2\,\mathrm{\Omega}_{\mathrm{(DM)},0}}{2\,k^2}\,(1+z)\,\delta_{\mathrm{(DM)}}(\vec{k},z)\,q(z) ,
    \end{align}
    with $\delta_{\mathrm{(DM)}}(\vec{k},z)=\delta^0_{\mathrm{(DM)}}(\vec{k})\,D_{\mathrm{(DM)}}(z)$. In general, the parameter  $q$ is a function of both redshift and scale. Since in the present work, we will investigate its behavior in terms of redshift, we calculate $q$ for a specific value of $k$. As long as we are interested in sub-Hubble scales, we choose $k=0.05\;\mathrm{Mpc}^{-1}$, which is also valid in linear regime. Considering Eqs. (\ref{eq30}), (\ref{eq31}), and (\ref{eq34}), we obtain
    \begin{align} \label{eq35}
    a^T_{\ell m}&=-\frac{3}{2\pi^2}\,H_0^2\,\mathrm{\Omega}_{\mathrm{(DM)},0}\,i^{\ell}\int\mathrm{d}^3 k\int^{z_i}_{0}\mathrm{d}z\,\delta^0_{\mathrm{(DM)}}(\vec{k})\,\frac{1}{k^2} \nonumber \\
    &\times j_{\ell}\big(kr(z)\big)Y^*_{\ell m}(\hat{k})\,\frac{\partial}{\partial z}\bigg((1+z)\,D_{\mathrm{(DM)}}(z)\,q(z)\bigg) .
    \end{align}
    Now, we look at galaxy density contrast defined as
    \begin{equation} \label{eq36}
    \delta_g(\vec{\theta})=\int{\mathrm{d}z\,b(z)\,\frac{\mathrm{d}N}{\mathrm{d}z}\,\delta_{\mathrm{(DM)}}(\vec{\theta},z)} ,
    \end{equation}
    in which $b(z)$ is the bias factor, and $\frac{\mathrm{d}N}{\mathrm{d}z}$ is a selection function which encapsulates the distribution of the galaxies observed by a survey, and normalized so that $\int{\mathrm{d}z\,\frac{\mathrm{d}N}{\mathrm{d}z}}=1$. Here we use \cite{PhysRevD.77.123520} for $\frac{\mathrm{d}N}{\mathrm{d}z}$. It is possible to derive $a^g_{\ell m}$ similarly as before,
    \begin{align} \label{eq38}
    a^g_{\ell m}&=\frac{1}{2\pi^2}\,i^{\ell}\,\int\mathrm{d}^3 k\,\int\mathrm{d}z\,j_{\ell}\big(kr(z)\big)Y^*_{\ell m}(\hat{k})\,b(z)\,\frac{\mathrm{d}N}{\mathrm{d}z} \nonumber \\
    &\times \delta^0_{\mathrm{(DM)}}(\vec{k})\,D_{\mathrm{(DM)}}(z)  .
    \end{align}
    Following Eqs. (\ref{eq35}) and (\ref{eq38}), the ISW-galaxy cross-correlation angular power spectrum $C^{gT}_{\ell}$ is written as
    \begin{align} \label{eq39}
    C^{gT}_{\ell} &\equiv \bigg<a^g_{\ell m}\,a^{T*}_{\ell' m'}\bigg> \nonumber \\
    &=\frac{2}{\pi}\int \mathrm{d}k\,k^2\,I_{\ell}^{ISW}(k)\,I_{\ell}^{g}(k)\,P^0_{\delta}(k) ,
    \end{align}
    where $I_{\ell}^{ISW}(k)$ and $I_{\ell}^{g}(k)$ are defined as
    \begin{align} \label{eq40}
    I_{\ell}^{ISW}(k)&=-\frac{3\,H_0^2\,\mathrm{\Omega}_{\mathrm{(DM)},0}}{k^2}\,\int^{z_i}_{0} \mathrm{d}z\,j_{\ell}\big(kr(z)\big) \nonumber \\
    &\times \frac{\partial}{\partial z}\bigg((1+z)\,D_{\mathrm{(DM)}}(z)\,q(z)\bigg) ,
    \end{align}
    \begin{equation} \label{eq41}
    I_{\ell}^{g}(k)=\int \mathrm{d}z\,j_{\ell}\big(kr(z)\big)\,b(z)\,\frac{\mathrm{d}N}{\mathrm{d}z}\,D_{\mathrm{(DM)}}(z) .
    \end{equation}
    The bias factor which relates density contrast of dark matter to galaxy distribution is in general a function of scale and redshift. Here we assume constant bias that depends on tracers. Redshift dependence of bias and its effect on ISW-galaxy cross-correlation have been considered in previous works \cite{1475-7516-2016-09-003}.
    \subsection{Galaxy power spectrum} \label{sec3.2}
    The measurement of galaxy power spectrum is a useful probe to set constraints on cosmological parameters. Galaxy power spectrum is related to matter power spectrum (in linear regime) as
    \begin{equation} \label{eq42}
    P_g(k)=b^2\,P_m(k) ,
    \end{equation}
    where $b$ is the bias factor. Although different cases of bias dependence on redshift have been studied before \cite{1475-7516-2014-01-019}, here we consider a constant bias model.
    \subsection{${{f \sigma_8}}$ measurements} \label{sec3.3}
    The large scale RSD measurements can be applied to study the growth of linear structures. The cosmological growth rate is defined as
    \begin{equation}
    f(z)=-\frac{1+z}{D(z)}\,\frac{\mathrm{d}D(z)}{\mathrm{d}z} .
    \end{equation}
    Early growth rate surveys have been analyzed to measure the parameter $\beta=f/b$, which depends on the galaxy bias. Looking for a bias-independent parameter, it has been found that $f \sigma_8$ measurements are able to discriminate between cosmological models \cite{doi:10.1111/j.1365-2966.2008.14211.x}. $\sigma_8$ has the following relation with the growth function:
    \begin{equation}
    \sigma_8(z)=\frac{\sigma_{8}(z=0)}{D(z=0)}\,D(z) .
    \end{equation}
    \subsection{Background probes} \label{sec3.4}
    Considering background data, we use the CMB data containing the physical matter density $\mathrm{\Omega}_{\mathrm{(M)},0}h^2$ and also $100\theta_*$, which is the ratio of the sound horizon to the angular diameter distance at recombination. 
    
    In the next section, numerical results from the above surveys for the interacting model are reported.
    \section{Analysis method and results} \label{sec4}
    In this section, we investigate the observational constraints on cosmological parameters of the interacting model. 
    Considering ISW-galaxy cross-correlation, we use the luminous red galaxies (LRGs) measurements, which are correlated with the CMB map of Wilkinson Microwave Anisotropy Probe (WMAP) third year data \cite{PhysRevD.77.123520}. Furthermore, we use galaxy power spectrum data \cite{Tegmark} (in which we have considered data points for the scales $k\lesssim0.09\;h\,\mathrm{Mpc}^{-1}$) and also the $f\sigma_8$ data displayed in Table \ref{tab1}, which are independent measurements of $f\sigma_8$ according to Ref. \cite{Nesseris}. In addition, we use the Planck 2015 data \cite{2016A&A...594A..13P} for background probes, which are $\mathrm{\Omega}_{\mathrm{(M)},0}h^2=0.1426$ and $100\theta_*=1.04105$. 
    \begin{table}[h!]
    	\centering
    	\caption{The data points of $f\sigma_8$ at different redshifts.}
    	\label{tab1}
    	\begin{tabular}{|c|c|c|}
    		\hline
    		$z$ &  $f\sigma_8$ & Ref. \\ \hline
    		$0.02$ & $0.428\pm0.0465$ & \cite{Huterer/0.02} \\
    		$0.1$ & $0.37\pm0.13$ & \cite{feix0.1} \\
    		$0.15$ & $0.49\pm0.145$ & \cite{Howlett/0.15} \\
    		$0.17$ & $0.51\pm0.06$ & \cite{song0.17} \\
    		$0.18$ & $0.36\pm0.09$ & \cite{Blake/0.18/0.38} \\
    		$0.38$ & $0.44\pm0.06$ & \cite{Blake/0.18/0.38} \\
    		$0.25$ & $0.3512\pm0.0583$ & \cite{samushia0.25/0.37} \\
    		$0.37$ & $0.4602\pm0.0378$ & \cite{samushia0.25/0.37} \\
    		$0.59$ & $0.488\pm0.06$ & \cite{chuang0.59} \\    			
    		$0.44$ & $0.413\pm0.08$ & \cite{blake0.44/0.6/0.73} \\
    		$0.60$ & $0.39\pm0.063$ & \cite{blake0.44/0.6/0.73} \\     		
    		$0.73$ & $0.437\pm0.072$ & \cite{blake0.44/0.6/0.73} \\
    		$0.86$ & $0.4\pm0.11$ & \cite{Pezzotta/0.86} \\
    		$1.4$ & $0.482\pm0.116$ & \cite{Okumura/1.4} \\ 		 
    		\hline
    	\end{tabular}
    \end{table}
    \subsection{Case study} \label{sec4.1}
    In this part, we consider different values of dark energy equation of state $w_\mathrm{(DE)}$ and coupling constant $\gamma$ in order to discuss the physical interpretation of the signal due to chosen parameters.

\begin{figure}[h!]
    	\centering
    	\includegraphics[width=8.5cm]{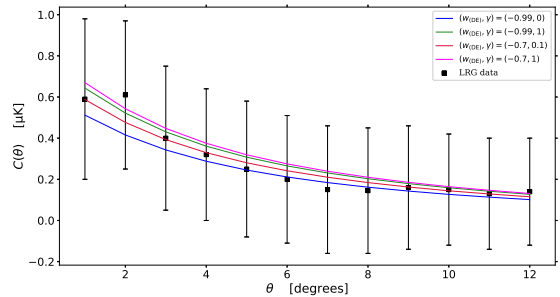}
    	\caption{The angular power spectrum of the ISW galaxy in term of the angle of separation for noninteracting model $(w_\mathrm{(DE)},\gamma)=(-0.99,0)$ and interacting model with different values of $(w_\mathrm{(DE)},\gamma)$, compared with the observational data. The bias parameter is set to a constant ($b=1.8$). The data points are taken from LRG sample.}
    	\label{fig4}
    \end{figure}

    Figure \ref{fig4} shows the angular power spectrum of ISW-galaxy cross-correlation for noninteracting and interacting models, using different values of ($w_\mathrm{(DE)},\gamma$).
    We use Eisenstein-Hu transfer function \cite{Eisenstein_1998} for computing the matter power spectrum. The data points in Fig. \ref{fig4} are from the CMB-LRG cross-correlation data, with LRG data extracted from the Sloan digital sky survey (SDSS) catalog \cite{doi:10.1111/j.1365-2966.2006.11263.x}. Since LRGs have a redshift distribution (with a mean redshift of $z\sim 0.5$) deeper than ordinary galaxies, they would be appropriate tracers of dark matter distribution (and hence, they have been used to find evidence for the ISW effect \cite{article,PhysRevD.72.043525}). In this analysis, we use the data points from \cite{PhysRevD.77.123520} which are processed from MegaZ LRG sample \cite{doi:10.1111/j.1365-2966.2006.11263.x,doi:10.1111/j.1365-2966.2006.11305.x} and contain 1.5 million objects from the SDSS DR6 selected with a neural network. Diagrams in Fig. \ref{fig4} are plotted with constant halo dark matter bias ($b=1.8$) \cite{PhysRevD.77.123520}.
        
\begin{figure}[h!]
		\centering
		\includegraphics[width=8.5cm]{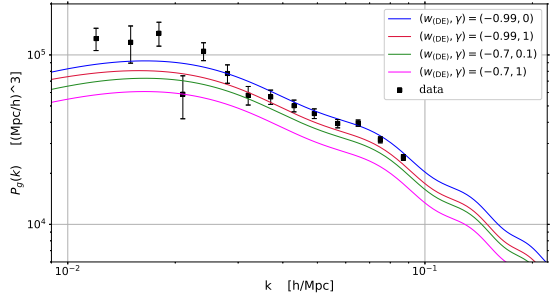}
		\caption{Galaxy power spectrum diagrams for noninteracting model $(w_\mathrm{(DE)},\gamma)=(-0.99,0)$ and interacting model with different values of $(w_\mathrm{(DE)},\gamma)$, compared with observational data from Ref. \cite{Tegmark}. The bias parameter is set to a constant ($b=1.9$).}
		\label{figpg}
	\end{figure}

\begin{figure}[h!]
		\centering
		\includegraphics[width=8.5cm]{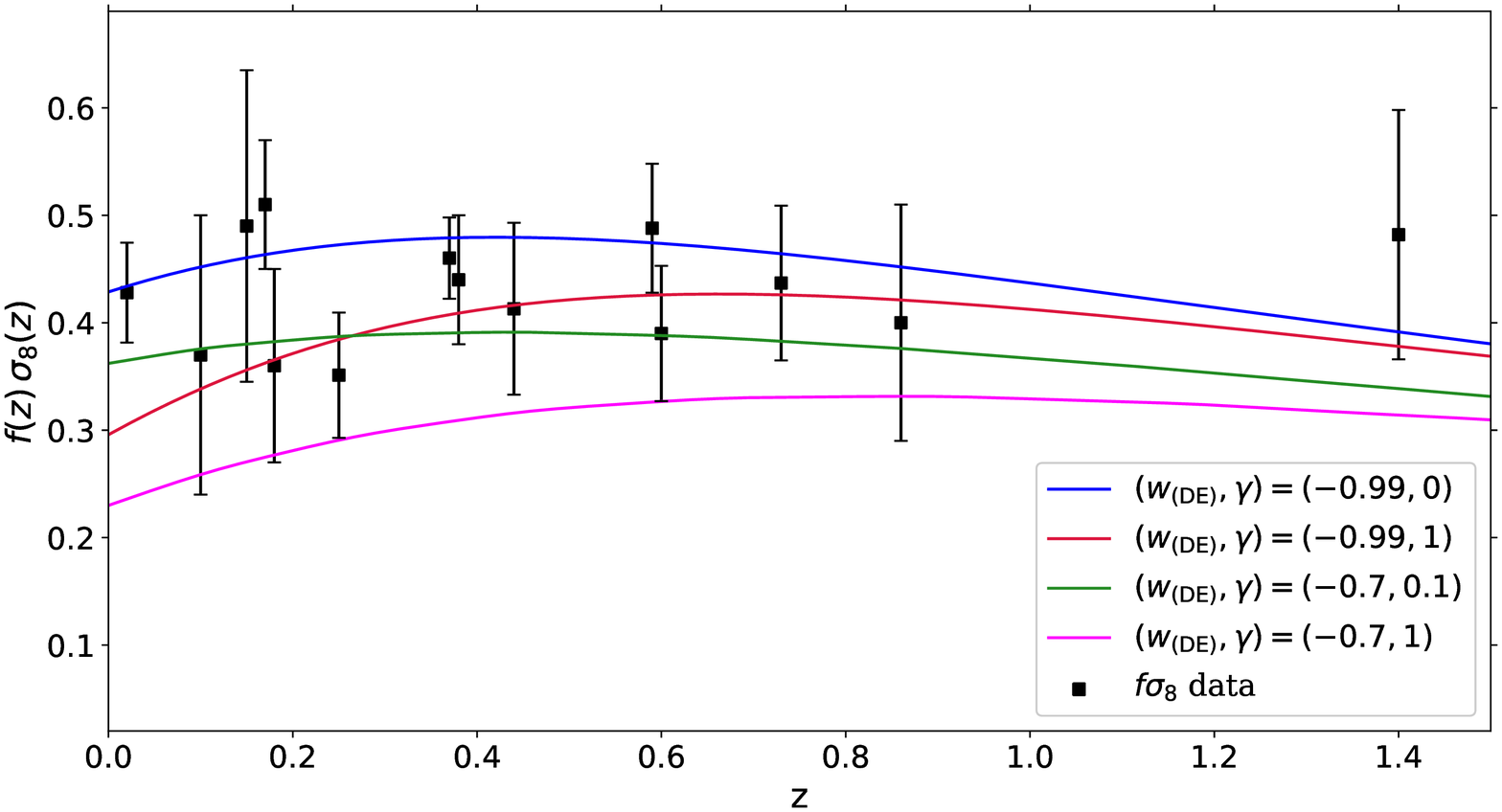}
		\caption{$f \sigma_8$ diagrams for noninteracting model $(w_\mathrm{(DE)},\gamma)=(-0.99,0)$ and interacting model with different values of $(w_\mathrm{(DE)},\gamma)$ compared with observational data displayed in Table \ref{tab1}.}
		\label{figfs}
	\end{figure}

    Figures \ref{figpg} and \ref{figfs} demonstrate galaxy power spectrum diagrams and $f \sigma_8$ diagrams, respectively. The galaxy power spectrum is calculated according to relation (\ref{eq42}) with $P_m(k)$ obtained from the CLASS code and considering constant bias ($b=1.9$). All chosen values of $(w_\mathrm{(DE)},\gamma)$ are compatible with ISW-galaxy cross-correlation data. However, it is found that interacting model with $(w_\mathrm{(DE)},\gamma)=(-0.7,1)$ is less favored by galaxy power spectrum data and $f\sigma_8$ data.	
	
\begin{figure}[h!]
		\centering
		\includegraphics[width=8.5cm]{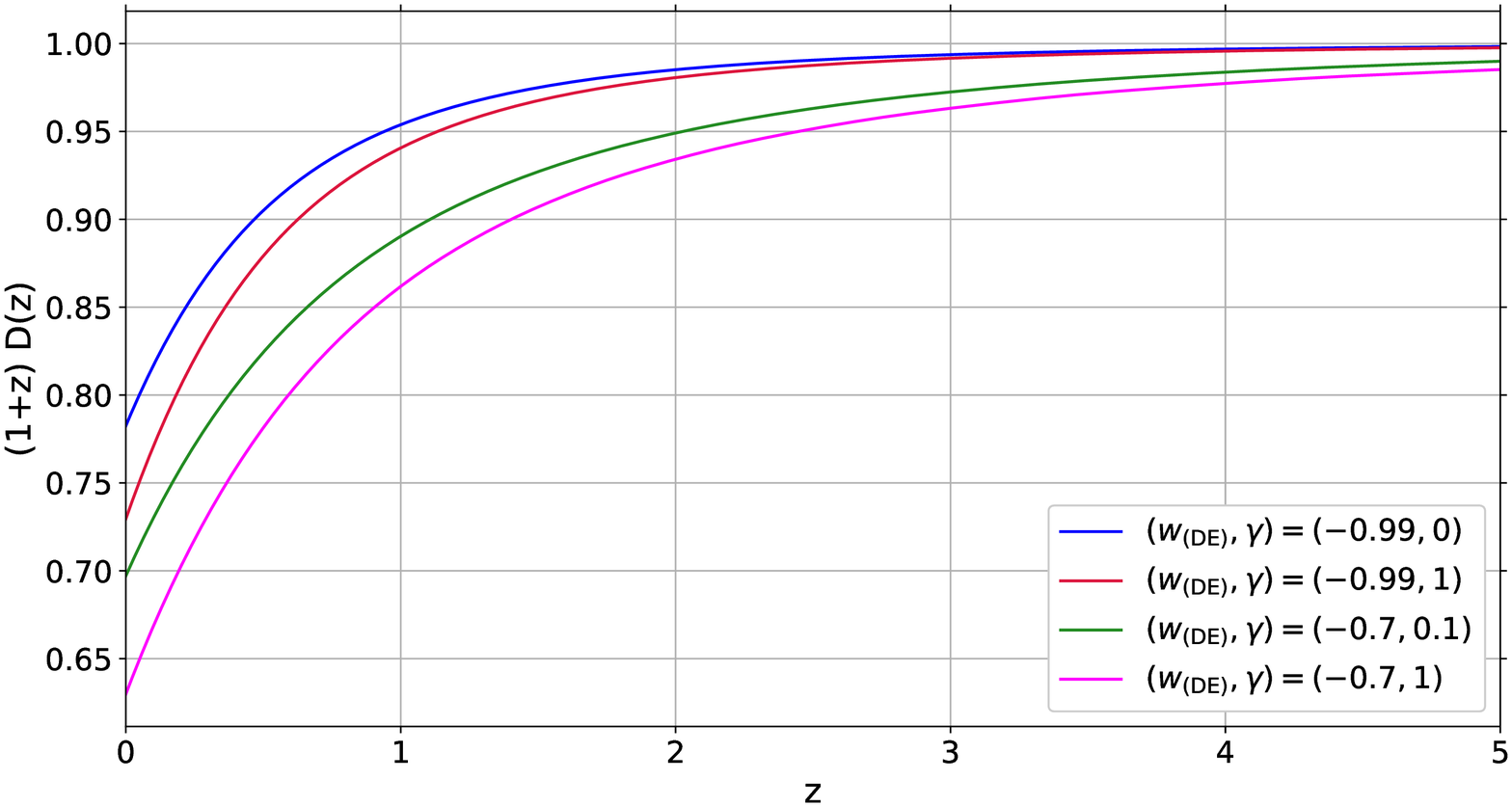}
		\caption{$(1+z)\,D(z)$ in terms of redshift for noninteracting model $(w_\mathrm{(DE)},\gamma)=(-0.99,0)$ and interacting model with different values of $(w_\mathrm{(DE)},\gamma)$.}
		\label{fig6}
	\end{figure}

	The growth function is illustrated in Fig. \ref{fig6} for noninteracting and interacting models. It can be seen that structure growth decreases in the presence of interaction. This feature has been already shown in matter power spectrum diagrams in Fig. \ref{fig3}.
	\subsection{Parameter estimation} \label{sec4.2}
	Here, we constrain the free parameters of the interacting model, using the combined ISW-galaxy cross-correlation (gT), galaxy power spectrum (P), $f\sigma_8$ (fs), $\mathrm{\Omega_{(M),0}}h^2$ (M), and $100\theta_*$ (th) dataset. We consider the following set of parameters in our analysis: 
	$\{\mathrm{\Omega_{(DM),0}}, w_\mathrm{(DE)},\gamma, H_0, \sigma_{8}(z=0)\}$.
	In numerical analysis, the total likelihood is defined as $\mathcal{L}_\mathrm{tot}\propto e^{-\chi^2_\mathrm{tot}/2}$, where $\chi^2_\mathrm{tot}$ is given by 
	\begin{equation}
	\chi^2_\mathrm{tot}=\chi^2_\mathrm{gT}+\chi^2_\mathrm{P}+\chi^2_\mathrm{fs}+\chi^2_\mathrm{M}+\chi^2_\mathrm{th} ,
	\end{equation}
	in which the terms on the right-hand side represent the $\chi^2$ values for observational probes.
	
	In our analysis, the matter power spectrum is computed with Eisenstein-Hu transfer function. 
	
	Table \ref{tab2} displays the best fit values with $1\sigma$ confidence levels of the cosmological parameters of our interacting dark energy model and also the ${\Lambda}$CDM model as a reference.
	\begin{table}[h!]
		\centering
		\caption{The best fit values of cosmological parameters with their $68\%$ confidence limits for interacting dark energy model and ${\Lambda}$CDM model using the combined gT+P+fs+M+th dataset.}
		\label{tab2}
		\begin{tabular}{|c|c|c|}
			\hline
			Parameter & Interacting model & ${\Lambda}$CDM model \\ \hline
			$\mathrm{\Omega_{(DM),0}}$ & $0.300\pm0.050$ & $0.300\pm0.040$ \\
			$w_\mathrm{(DE)}$\footnote{Dark energy equation of state has an upper limit only, caused by the chosen prior range on this parameter according to preliminary numerical works.} & $-0.990+0.16$ & --- \\
			$\gamma$ & $0.150^{+9.8}_{-0.15}$ & --- \\
			$H_0\,[\mathrm{\frac{km}{s\,Mpc}}]$ & $67.5\pm3.8$ & $70.0\pm3.0$ \\
			$\sigma_{8}(z=0)$ & $0.700\pm0.28$ & $0.800\pm0.18$ \\
			\hline
		\end{tabular}
	\end{table}
	According to the observational data, the interacting model prefers a lower value of $\sigma_8$ compared to ${\Lambda}$CDM model. Consequently, the interacting model is capable of alleviating the $\sigma_8$ tension. This result is according to the fact that interaction in the dark sector prevents dark matter from clustering [as understood from Eq. (\ref{eqdm})] and consequently suppresses structure growth, which yields to lower values of $\sigma_8$. Additionally, according to preliminary numerical works, the perturbation equations of interacting model might diverge by choosing $w_\mathrm{(DE)}<-1$, so we focused on quintessential dark energy and considered the prior range $[-0.99,-0.5]$ for dark energy equation of state. Hence, there is only an upper limit on $w_\mathrm{(DE)}$ as reported in Table \ref{tab2}.
	
	As the final point in this section, we use the Akaike information criterion (AIC), which is a means for model selection \cite{Akaike1974AIC,burnham}, in order to obtain the goodness of fit of our models to observational data. AIC is defined as \cite{Akaike1974AIC}
	\begin{equation} 
		\mathrm{AIC}=-2\,\ln{\mathcal{L}_{\mathrm{max}}}+2\,K ,
	\end{equation}
	in which $\mathcal{L}_{\mathrm{max}}$ is the maximum likelihood function, and $K$ is the number of free parameters. Comparing models, the one that minimizes AIC can be considered as the best model. According to the likelihood for the present model, we obtain
	\begin{align*}
	& \mathrm{AIC}=68.33 \;\; \text{(interacting model)} , \\
	& \mathrm{AIC}=70.5 \;\; \text{($\Lambda$CDM model)} ,
	\end{align*}	
	and so $\Delta\mathrm{AIC}=2.17$. Hence, it can be concluded that the observational data favor the interacting model as well as the $\Lambda$CDM model.
	\section{Forecast analysis} \label{sec5}
	In this section, we forecast constraints on cosmological parameters of the interacting dark energy model. Accordingly, we consider both spectroscopic and photometric redshift surveys, i.e., Euclid-like \cite{euclid1,euclid2} and  LSST\footnote{Large Synoptic Survey Telescope.}-like \cite{lsst} surveys, respectively.
	
	In order to estimate the accuracy of the interacting model parameters, we apply the Fisher matrix formalism \cite{fisher}. The Fisher matrix is defined as
	\begin{equation} \label{f1}
	F_{xy}=-\big<\frac{\partial^2\ln{\mathcal{L}}}{\partial p_x\,\partial p_y}\big> ,
	\end{equation}
	in which $\mathcal{L}$ is the likelihood function, and $p$ is a parameter of the cosmological model. According to the Cramer-Rao inequality, the Fisher matrix approach provides us with the best estimate of the model parameter errors. In other words, for a model likelihood with a Gaussian distribution, the inverse of Fisher matrix is the covariance matrix of the parameters which approximates the parameter errors. So, in a simultaneous estimation of all parameters, the $1\sigma$ error on parameter $p_x$ is $\sqrt{{(F^{-1})}_{xx}}$.
	
	In our analysis, the likelihood function is the galaxy power spectrum $P_g$, which can be written as \cite{doi:10.1093/mnras/227.1.1,1475-7516-2012-04-005}
	\begin{equation}
	P_g(k,\mu;z)=(b+f\,\mu^2)^2\,D(z)^2\,P_m(k)\,e^{-k^2\mu^2\sigma_r^2} ,
	\end{equation}
	where $b$ is the bias factor, $D(z)$ is the growth function, $f$ is the growth rate, $P_m(k)$ is the matter power spectrum at $z=0$, $\mu$ is the cosine of the angle between the line of sight and wave number, and
	\begin{equation}
	\sigma_r=\frac{c\,\sigma_z}{H(z)} ,
	\end{equation}
	with $\sigma_z$ the absolute error on redshift measurement. The matter power spectrum is defined as \cite{amendola_tsujikawa}
	\begin{equation}
	P_m(k)=\frac{2\pi^2 \delta_H^2}{\mathrm{\Omega}_\mathrm{(M),0}^2}\,a_0^2\,{(\frac{k}{H_0})}^{n_s}\,H_0^{-3}\,T(k)^2\,D(a_0)^2 ,
	\end{equation}
	in which $\delta_H^2$ is the amplitude of gravitational potential, $n_s$ is the spectral index, and $T(k)$ is the Eisenstein-Hu transfer function.
	
	The Fisher matrix for a redshift bin $z_i$ takes the form \cite{PhysRevLett.79.3806}
	\begin{align}
	F_{xy}(z_i)&=\frac{1}{8 \pi^2}\,\int^1_{-1} \mathrm{d}\mu \, \int^{k_\mathrm{max}}_{k_\mathrm{min}} \mathrm{d}k \, k^2 \, \frac{\partial \ln{P_g(k,\mu;z_i)}}{\partial p_x} \nonumber \\
	&\times \frac{\partial \ln{P_g(k,\mu;z_i)}}{\partial p_y}\,(\frac{\bar{n}_i\,P_g(k,\mu;z_i)}{\bar{n}_i\,P_g(k,\mu;z_i)+1})^2\,V_i ,
	\end{align}
	where $\bar{n}$ is the mean number density of galaxies, and $V$ is the survey volume.
	The volume of a redshift bin is given by
	\begin{equation}
	V_i=\frac{4\pi}{3}\,f_\mathrm{sky}\,\big(d_c^3(z_\mathrm{max})-d_c^3(z_\mathrm{min})\big) ,
	\end{equation}
	in which $f_\mathrm{sky}$ is the fractional sky coverage of the survey, and $d_c$ is the comoving distance to redshift $z$,
	\begin{equation}
	d_c(z)=\int^z_0 \frac{c}{H(z)}\,\mathrm{d}z .
	\end{equation}
	And also, the mean number density of a redshift bin is defined by
	\begin{equation}
	\bar{n}_i=\frac{4\pi}{V_i}\,f_\mathrm{sky}\,\int^{z_\mathrm{max}}_{z_\mathrm{min}} \mathrm{d}z \frac{\mathrm{d}N}{\mathrm{d}z}(z) ,
	\end{equation}
	with $\frac{\mathrm{d}N}{\mathrm{d}z}(z)$ the surface number density of survey. $k_\mathrm{min}$ for each redshift bin is $k_\mathrm{min}=2\pi\,{(\frac{3\,V_i}{4\pi})}^{-1/3}$, and we define $k_\mathrm{max}$ as $k_\mathrm{max}=\frac{\pi}{2\,R}$, in which $\sigma_R(z)=\frac{1}{2}$, in order to exclude information from nonlinear regime.
	
	In the following, we explain specifications of Euclid-like and LSST-like surveys. For the Euclid-like survey, we consider the redshift range $0.7<z<2.1$, which is divided into seven redshift bins with the mean redshifts: $0.8,\,1.0,\,1.2,\,1.4,\,1.6,\,1.8,\,2.0$. We assume $f_\mathrm{sky}=0.36$, $\sigma_z=0.001\,(1+z)$, and the galaxy bias $b(z)=\sqrt{1+z}$. The surface number density $\frac{\mathrm{d}N}{\mathrm{d}z}(z)$ has been chosen from Ref. \cite{doi:10.1111/j.1365-2966.2009.15977.x}, considering the flux limit be equal to $3\times 10^{-16}\,\frac{\mathrm{erg}}{\mathrm{cm^2\,s}}$, with an efficiency of $30\%$.
	
	For the LSST-like survey, we follow Ref. \cite{1475-7516-2016-11-014}. Hence, we assume seven redshift bins in the redshift range $0.2<z<3.0$, with the mean redshifts $0.31,\,0.55,\,0.84,\,1.18,\,1.59,\,2.08,\,2.67$. We also consider $f_\mathrm{sky}=0.58$, $\sigma_z=0.04\,(1+z)$, and $b(z)=1+0.84z$. The surface number density takes the form \cite{0004-637X-644-2-663}
	\begin{equation}
	\frac{\mathrm{d}N}{\mathrm{d}z}(z)=640\,z^2\,e^{-z/0.35} \; \mathrm{arcmin}^{-2} .
	\end{equation}
	
	As fiducial cosmology for our Fisher forecast, we consider $(\mathrm{\Omega}_\mathrm{(DM),0},H_0,w_{\mathrm{(DE)}},\gamma,\sigma_8(z=0),n_s)=(0.300,67.5\,\mathrm{\frac{km}{s\,Mpc}},-0.990,\,0.150,0.700,0.9655)$, in which the value of $n_s$ is according to the Planck 2015 data \cite{2016A&A...594A..13P}, and the values of other parameters are the best fit values obtained from the observational data. 
	
\begin{figure*}
		\centering
		\includegraphics[width=8.5cm]{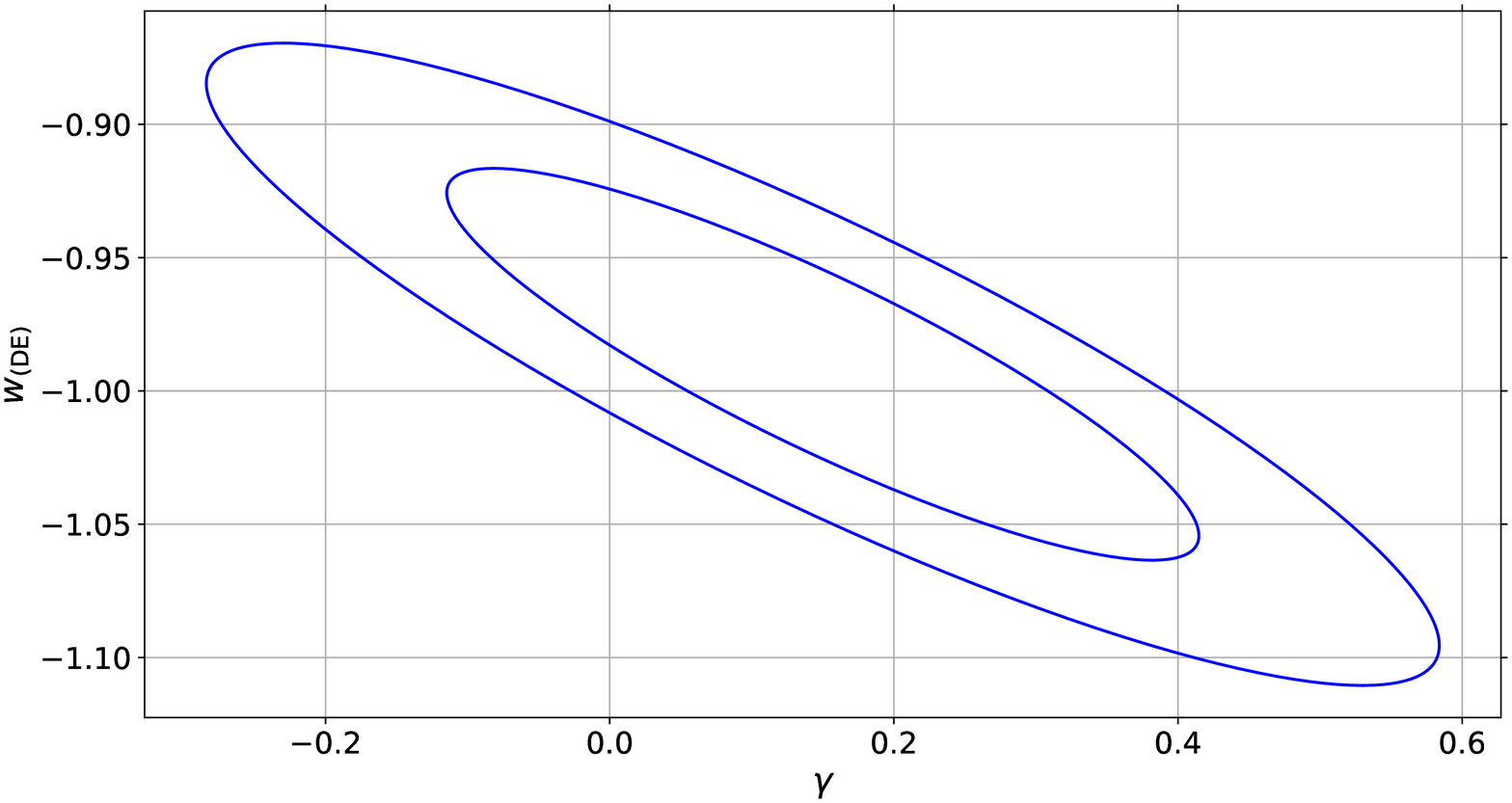}
		\includegraphics[width=8.5cm]{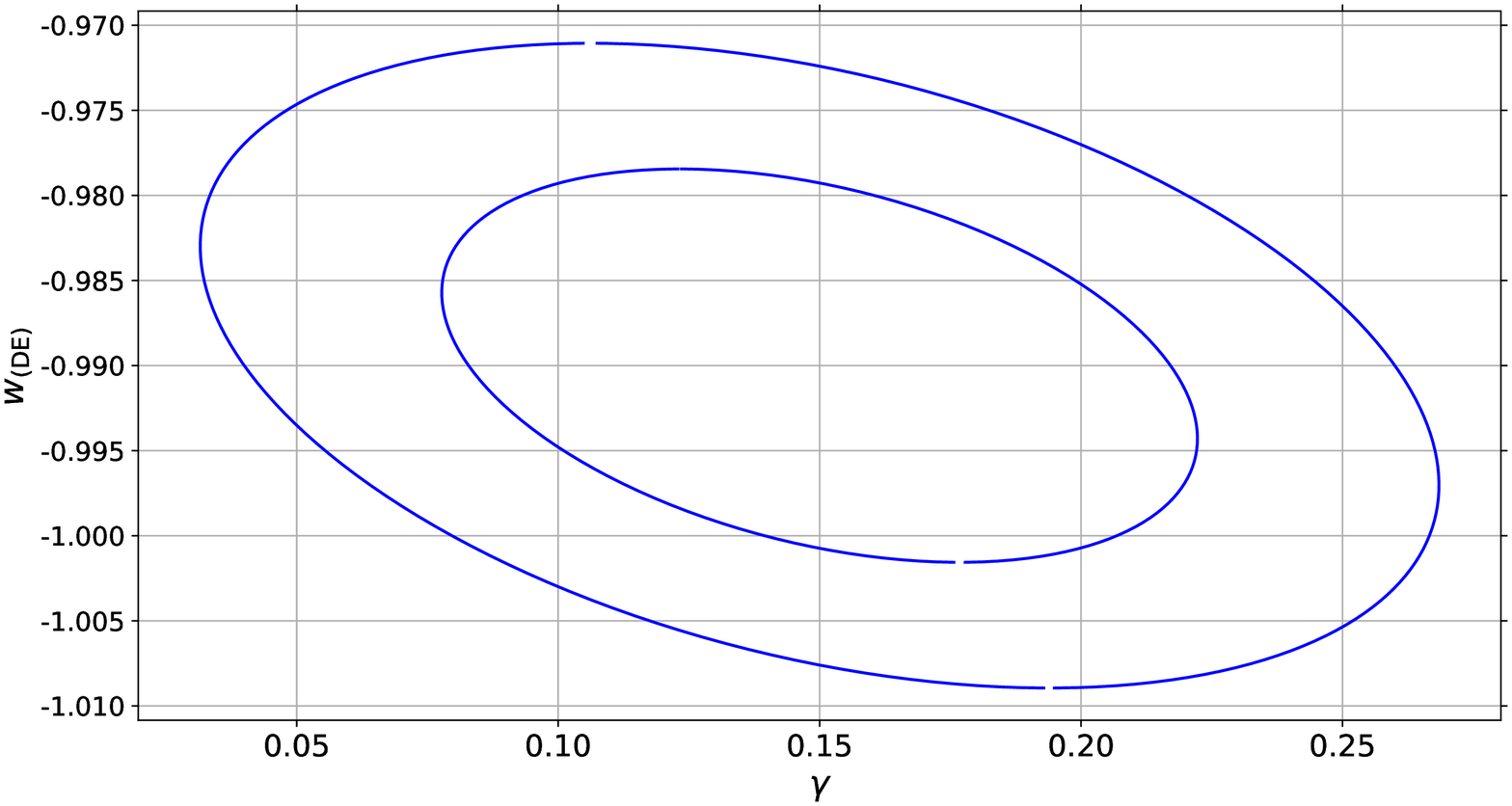}
		\caption{Forecasted $1\sigma$ and $2\sigma$ contours on the $w_\mathrm{(DE)}$-$\gamma$ plane, considering Euclid-like survey (left plot) and LSST-like survey (right plot).}
		\label{figf1}
	\end{figure*}

	Figure \ref{figf1} shows the predicted $1\sigma$ and $2\sigma$ contours for Euclid-like and LSST-like surveys.
	The LSST-like survey, which has larger sky coverage, places tighter constraints on parameters of the interacting dark energy model.
	\section{Conclusions} \label{sec6}
    The standard cosmological model (known as ${\Lambda}$CDM model) provides a satisfactory description of the large scale structure evolution of Universe. However, there are fundamental issues such as the unknown nature of dark energy and dark matter and also the cosmological constant problem. Furthermore, there is a list of observations which introduces a slightly $2-3\sigma$ tension with the standard model. These tensions could be due to statistical errors or observational inaccuracies. However, some of them may indicate new physics beyond the standard cosmological model. One of the interesting tensions to investigate is the discrepancy of matter density perturbations power due to early and late-time observations. It seems that the CMB data predict more power in late times than the one obtained from large scale structure observations such as cluster count, weak lensing, and the redshift space distortion. This tension is usually formulated by $\sigma_8$ or $f\sigma_8$ observations. Accordingly, this tension can be a hint of new physics in the dark sector. Regarding this, we have investigated the capability of the interacting dark energy model described in Eqs. (\ref{eq20})--(\ref{eq25}) to relieve the $\sigma_8$ tension. In this direction, we use the ISW-galaxy cross-correlation, galaxy power spectrum, $f\sigma_8$, and CMB data to study the interacting model. We should note that the proposed model introduces new interaction in perturbative level. This can be an interesting idea for standard model extensions which can be studied as a potential proposal to resolve the tensions in $\sigma_8$ observations.  All of these must be considered along with the fact that we have taken a constant dark energy equation of state and also a constant bias parameter. Moreover, we have employed Fisher matrix approach to perform forecasts for the parameters of the interacting dark energy model from Euclid-like and LSST-like surveys.
	
	Considering numerical results, the interacting dark energy model is supported by observational data as well as the $\Lambda$CDM model, while having the advantage to alleviate the $\sigma_8$ tension. As shown here, this better result is due to the fact that interaction between dark matter and dark energy suppresses structure growth in Universe which lowers the best fit value for $\sigma_8$. Also, regarding the Fisher forecast results, the LSST-like survey could place better constraints on the interacting model parameters and also show deviations from the standard cosmological model. 
	\\	
	\begin{acknowledgments}
	The authors wish to thank especially Shant Baghram for his manuscript reading and valuable comments.
	\end{acknowledgments}
	\interlinepenalty=10000
	\bibliography{paper}
\end{document}